\documentclass[12pt]{iopart}
\usepackage{graphicx}
\usepackage{dcolumn}
\usepackage{bm}
\usepackage{tabularx}
\usepackage{ulem}
\newcolumntype{M}{>{\centering\arraybackslash}m{1.85cm}}
\usepackage[export]{adjustbox}
\usepackage{float}
\usepackage{braket}
\usepackage{makecell}
\usepackage{lipsum}
\usepackage{longtable}
\graphicspath{{figure/}}
\usepackage{xcolor}   

\usepackage{hyperref}

\begin{document}

\title{Large-scale shell-model study of 2$\nu$ECEC process in $^{78}$Kr}

\author{Deepak Patel, Praveen C. Srivastava}
\address{Department of Physics, Indian Institute of Technology Roorkee, Roorkee
	247 667, India}

\vspace{10pt}

%

\begin{abstract}

In this work, we present the systematic study of $2\nu$ECEC process in the $^{78}$Kr using large-scale shell-model calculations with the GWBXG effective interaction. We first validate the efficiency of the utilized interaction by comparing the theoretical low-lying energy spectra, the kinematic moment of inertia, and reduced transition probabilities with the experimental data for both the parent and grand-daughter nuclei $^{78}$Kr and $^{78}$Se, respectively. Additionally, we examine the shell-model level densities of the $1^+$ states in the intermediate nucleus $^{78}$Br, comparing them with the predictions from the Back-shifted Fermi gas model. We analyze the variation of cumulative nuclear matrix elements (NMEs) for the $2\nu$ECEC process in $^{78}$Kr as a function of $1^+$ state energies in the intermediate nucleus $^{78}$Br up to the saturation level. Our estimated half-life for $^{78}$Kr, extracted from the shell-model predicted NMEs, shows good agreement with the experimental value. The Gamow-Teller transitions from the lowest $1^+$ state of $^{78}$Br via both the EC$+\beta^+$ and $\beta^-$-channels are also discussed.

\end{abstract}

%
%
%
%
%

\section{Introduction}

Double electron capture (ECEC) is one of the rarest weak interaction processes and has been a subject of significant interest in the field of nuclear and particle physics over the last few decades \cite{Gavrilyuk, Ratkevich,  Suhonen5, Pirinen, Coello, Nitescu1, Mishra, Aprile, Aprile1, Aprile2, Nitescu, Patel, Abe, Aalbers, Meshik, Agostini, Laubenstein}. This process involves the simultaneous capture of two electrons by a nucleus and can occur in two distinct modes: two-neutrino double electron capture (2$\nu$ECEC) and neutrinoless double electron capture (0$\nu$ECEC) \cite{Aprile}. These modes are represented by the reactions $(Z, A) + 2e^- \rightarrow (Z-2, A) + 2\nu_e$ for 2$\nu$ECEC and $(Z, A) + 2e^- \rightarrow (Z-2, A)$ for 0$\nu$ECEC \cite{Abe}. The ECEC process primarily takes place in those nuclei where single $\beta^+/$EC process is either energetically forbidden or significantly suppressed. The 2$\nu$ECEC process adheres to all conservation laws of the standard model. However, the 0$\nu$ECEC, which is not observed yet, violates lepton number conservation, providing key insights into the Majorana nature of neutrinos and physics beyond the standard model \cite{Blaum, Bustabad, Lema}. In this study, we consider only the $2\nu$ECEC process, which is essential for probing nuclear structure and accessing the effective value of weak axial-vector coupling constant.

The ECEC process is the less explored counterpart of the double beta-minus decay due to its longer half-lives and lower Q-values, which pose significant challenges for experimental detection \cite{Meshik}. Despite this, the 2$\nu$ECEC process has been observed in three nuclei $^{78}$Kr \cite{Gavrilyuk, Ratkevich}, $^{124}$Xe \cite{Aprile1, Aprile2}, and $^{130}$Ba \cite{Meshik, Pujol}. There are other prominent candidates of the 2$\nu$ECEC process, such as $^{126}$Xe and $^{132}$Ba, but only lower limits of their half-lives can be observed so far \cite{Abe, Meshik, Akerib}. In these circumstances, the theoretical and experimental investigation of the 2$\nu$ECEC process in the above nuclei becomes crucial. For $^{78}$Kr, the well-known candidate of 2$\nu$ECEC process, the half-life has been experimentally measured in two different experiments as $[9.2^{+5.5}_{-2.6}(\rm stat)\pm1.3(\rm syst)]\times 10^{21}$ yr using a
copper low-background proportional counter \cite{Gavrilyuk}, and $[1.9^{+1.3}_{-0.7}(\rm stat)\pm0.3(\rm syst)]\times 10^{22}$ yr using the time-resolving current pulse from the large low-background proportional counter (LPC) \cite{Ratkevich}. Therefore, it is essential to assess the effectiveness of various nuclear models in calculating the 2$\nu$ECEC nuclear matrix element (NME) for $^{78}$Kr, which plays an important role in determining the accurate half-life.

In recent decades, various theoretical approaches, including the quasiparticle random-phase approximation (QRPA), the interacting boson model (IBM), the Hartree-Fock-Bogoliubov model (HFB), and the nuclear shell model (NSM) have been employed to study $\beta^-\beta^-$, $\beta^+\beta^+$, $\beta^+$EC, and ECEC processes \cite{Suhonen1, Pacearescu, Yousef, Barea, Nomura, Raina, Kostensalo1, Horoi, Jia}. The nuclear shell model is a widely adopted theoretical framework for studying the nuclear structure and decay processes. Several calculations have been performed to determine the NMEs for $\beta^-\beta^-$ decay using the shell-model \cite{Patel, Kostensalo1, Horoi, Kostensalo, Caurier, Brown, Coraggio1, Shimizu1}, while the 2$\nu$ECEC process has been less explored within this framework. With new advancements in computational capabilities, it is now feasible to accurately estimate the 2$\nu$ECEC NME by including a large number of $1^+$ states in the intermediate nucleus for this decay process. For a precise prediction of the NME, it is crucial to ensure that the employed model space can sustain the level density of $1^+$ states in the intermediate nucleus across the relevant energy range. In this context, the efficiency of the effective interaction and model space can be evaluated by comparing the level density of these $1^+$ states with that predicted by simple nuclear level density models, such as the Back-shifted Fermi gas model (BFM) \cite{Bethe, Gilbert}.

Motivated by the recent experimental data on 2$\nu$ECEC studies \cite{Gavrilyuk, Ratkevich} and  
advancements in the computational techniques, we have conducted large-scale shell-model calculations to investigate the 2$\nu$ECEC process in $^{78}$Kr. The primary goal of this study is to determine a precise value for the 2$\nu$ECEC NME in $^{78}$Kr and provide an accurate prediction of the half-life close to the experimental value. Additionally, we aim to examine the level density of $1^+$ states in the intermediate nucleus $^{78}$Br and explore the variation in cumulative NME with respect to the energy of these $1^+$ states.

This paper is organized as follows: Section \ref{section2} provides a brief overview of the theoretical framework used to calculate the 2$\nu$ECEC NME and half-life, along with the effective shell-model interaction employed in this work. In Section \ref{section3}, we present the spectroscopic properties of the parent ($^{78}$Kr) and grand-daughter ($^{78}$Se) nuclei, the level density of $1^+$ states in the $^{78}$Br, the variation of cumulative NME, the estimated half-life of $^{78}$Kr for the 2$\nu$ECEC process using the final cumulative NME, the role of Gamow-Teller (GT) transitions from the lowest $1^+$ state of $^{78}$Br via both the EC$+\beta^+$ and $\beta^-$-channels and their partial half-lives. Finally, Section \ref{section4} summarizes the outcomes and draws the conclusions.

\section{Theoretical Framework} \label{section2}

\subsection{Half-life}

The half-life for the 2$\nu$ECEC process can be given as follows

\begin{equation}
T_{1/2}^{2\nu}=\frac{1}{G_{2\nu}^{\rm ECEC}(g_A^{\rm eff})^4|M_{2\nu}|^2},
\label{eq1}
\end{equation}

where $G_{2\nu}^{\rm ECEC}$ \cite{Nitescu, Kotila} represents the phase-space factor and $g_A^{\rm eff}$ corresponds to effective axial-vector coupling constant \cite{Suhonen4}. The NME $M_{2\nu}$ for $2\nu$ECEC process is given by \cite{Suhonen2, Suhonen3}

\begin{equation}
M_{2\nu}=\sum_{k}\frac{\langle 0_{\rm g.s.}^{(f)}||\sigma\tau^{+}||1_k^+\rangle \langle 1_k^+||\sigma\tau^{+}||0_{\rm g.s.}^{(i)}\rangle}{[\frac{1}{2}Q_{\rm ECEC}+E(1^+_k)-M_i]/m_e+1}.
\label{eq2}
\end{equation}

Here, $E(1^+_k)-M_i$ represents the energy difference between the $k^{th}$ intermediate $1^+$ state and the ground state (g.s.) of the initial nucleus; $m_e$ denotes the rest mass of the electron; $0_{\rm g.s.}^{(i)}$($0_{\rm g.s.}^{(f)}$) corresponds the ground state of the initial (final) nuclei; $\sigma$ stands for the pauli matrix; $\tau^+$ represents the isospin raising operator. $Q_{\rm ECEC}$ ($Q$-value) signifies the energy released during the process. $\langle 0_{\rm g.s.}^{(f)}||\sigma\tau^{+}||1_k^+\rangle$ (or $\langle 1_k^+||\sigma\tau^{+}||0_{\rm g.s.}^{(i)}\rangle$) refers the reduced GT matrix element. It is important to note that within the shell-model framework, the effective axial-vector coupling constant, $g_A^{\rm eff}$, in Eq. (\ref{eq1}) arises from the effective renormalization of the spin-isospin operator $\sigma\tau^+$ in the NME of Eq. (\ref{eq2}). By factoring $g_A^{\rm eff}$ outside the squared NME in Eq. (\ref{eq1}), the $M_{2\nu}$ in Eq. (\ref{eq2}) becomes independent of the effective axial coupling.

\subsection{Model space and Hamiltonian}

The shell-model Hamiltonian can be represented as follows

\begin{equation}
		H=\sum_{\alpha}\varepsilon_{\alpha}{\hat N}_{\alpha}+\frac{1}{4}\sum_{\alpha\beta\delta\gamma J}\langle j_{\alpha}j_{\beta}|V|j_{\gamma}j_{\delta}\rangle_{J} \times O^{\dag}_{J;j_{\alpha}j_{\beta}} 
		O_{J;j_{\delta}j_{\gamma}},
	\end{equation}
where, $\alpha=\{nlj\}$ stands for the single-particle orbitals and $\varepsilon_{\alpha}$ denote the corresponding single-particle energies. $\hat{N}_{\alpha}$  corresponds the particle number operator. $\langle j_{\alpha}j_{\beta}|V|j_{\gamma}j_{\delta}\rangle_{J}$ refers two-body matrix element. $O_{J}^{\dag}$, and $O_{J}$ are the fermion pair creation and annihilation operators, respectively.

The GWBXG effective interaction \cite{Hosaka, Ji, Gloeckner, Serduke} is utilized for calculating the reduced GT matrix elements. The mean-field part of the shell-model Hamiltonian corresponding to the GWBXG interaction consists of $0f_{5/2}1p0g_{9/2}$ proton and $1p_{1/2}0g1d2s$ neutron orbitals. Previously, this interaction has been employed to investigate the structure of different nuclei \cite{Dey, Kumbartzki, Stuchbery}. We have also used this interaction in our recent study on $2\nu\beta^-\beta^-$-decay \cite{Patel2}. It is not possible to perform our calculations in the full model space due to the huge shell-model dimensions. Thus,  we have employed truncations by completely restricting neutron excitation across $N=50$ shell and allowed the neutrons to fill in the $1p_{1/2}$ and $0g_{9/2}$ orbitals only. To ensure the saturation level in the cumulative NMEs, we have considered 5000 $1^+$ states in the intermediate nucleus $^{78}$Br for GT transitions. The shell-model code NuShellX \cite{Nushellx} has been utilized in the diagonalization of the shell-model Hamiltonian matrices.

\section{Results and Discussions} \label{section3}

In this section, we present the shell-model calculated results for the $2\nu$ECEC process in $^{78}$Kr. We first compare the calculated spectroscopic properties, including the low-lying energy spectra, kinematic moment of inertia, and reduced quadrupole transition probabilities, in the $^{78}$Kr and the $^{78}$Se nuclei with the corresponding experimental data. Next, we discuss the shell-model level densities of the $1^+$ states in the $^{78}$Br and compare them with the predictions from the BFM. We also analyze the variation of the cumulative NME for the $2\nu$ECEC process in $^{78}$Kr as a function of the $1^+$ state energies in $^{78}$Br and estimate the half-life using the final cumulative NME. Finally, we briefly discuss the GT transitions from the lowest $1^+$ state of $^{78}$Br via both the EC+$\beta^+$ and $\beta^-$-channels.

\begin{figure}
	\centering
	\includegraphics[width=15cm]{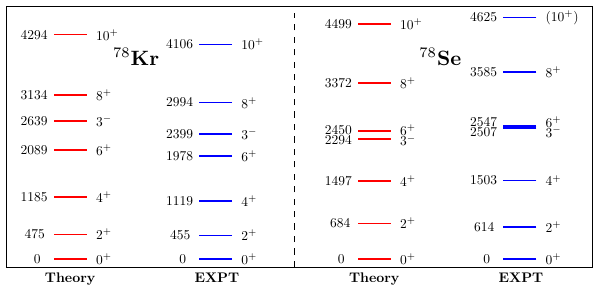}
	\caption{\label{energy_spectra} Comparison between the theoretical and experimental \cite{NNDC} energy levels in the parent ($^{78}$Kr) and grand-daughter ($^{78}$Se) nuclei.}
\end{figure}

\subsection{Spectroscopic properties}

\begin{figure}
	\centering
	\includegraphics[width=15.4cm]{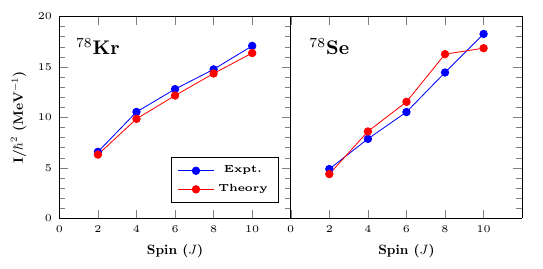}
	\caption{\label{kinematic_inertia} Comparison of the theoretical and experimental kinematic moment of inertia with respect to yrast positive-parity spins.}
\end{figure}

It is required to test the efficiency of the present effective interaction before calculating 2$\nu$ECEC NME in $^{78}$Kr. For this, we have compared the shell-model calculated low-lying energy spectra, kinematic moment of inertia (see Figs. \ref{energy_spectra} and \ref{kinematic_inertia}) and reduced quadrupole transition probabilities between the yrast states (see Table \ref{BE2}) in the $^{78}$Kr and $^{78}$Se with the experimental data.

\begin{table}
		\centering
		\caption{Comparison of the theoretical and experimental $B(E2)$ strengths (in Weisskopf unit (W.u.)) \cite{NNDC_NUDAT} in the parent and grand-daughter nuclei of $2\nu$ECEC process. The effective charges are taken as $e_\pi=1.7e,e_\nu=1.1e$ \cite{Boelaert}.}
  
			\begin{tabular}{cccccc}
				
				\hline
				
				Isotope	&	$J_i^{\pi} \rightarrow J_f^{\pi}$   & Theory  &	Expt. 	 \\
				
				\hline 
				
				$^{78}$Kr	&	$2^{+}_{1}$$\rightarrow$$0^{+}_{1}$   & 43.3 & 67.9(22) \\
				&	$4^{+}_{1}$$\rightarrow$$2^{+}_{1}$   & 62.5 & 88(5) \\
                    &	$6^{+}_{1}$$\rightarrow$$4^{+}_{1}$   & 68.2 & 94(11) \\
                    &	$8^{+}_{1}$$\rightarrow$$6^{+}_{1}$   & 67.5 & $\approx$85 \\
                    &	$10^{+}_{1}$$\rightarrow$$8^{+}_{1}$   & 63.8 & 80(12) \\

                    $^{78}$Se	&	$2^{+}_{1}$$\rightarrow$$0^{+}_{1}$   & 28.3 & 33.5(8) \\
                    &	$4^{+}_{1}$$\rightarrow$$2^{+}_{1}$   & 38.6 & 49.5(24) \\
                    &	$6^{+}_{1}$$\rightarrow$$4^{+}_{1}$   & 38.4 & 47(14) \\
                    &	$8^{+}_{1}$$\rightarrow$$6^{+}_{1}$   & 33.9 & 56(19)  \\
                    &	$10^{+}_{1}$$\rightarrow$$8^{+}_{1}$   & 42.7 & - \\

                \hline
				
			\end{tabular}
		\label{BE2}
	\end{table}

  It is clear from Fig. \ref{energy_spectra} that the present effective interaction successfully reproduces the low-lying yrast states in the $^{78}$Kr and $^{78}$Se. Shell-model calculations predict that the $0^+-6^+$ states in $^{78}$Kr are primarily dominated by the $\pi(f_{5/2}^4p_{3/2}^2g_{9/2}^2)\otimes \nu(g_{9/2}^4)$ and $\pi(f_{5/2}^2p_{3/2}^4g_{9/2}^2)\otimes \nu(g_{9/2}^4)$ configurations. A smaller but notable contribution from the $\pi(f_{5/2}^3p_{3/2}^3g_{9/2}^2)\otimes \nu(g_{9/2}^4)$ configuration also comes in the above states, which becomes the most dominating component in the $8^+-10^+$ states. The dominance of the $\pi(g_{9/2})$ orbital in the low-lying structure of $^{78}$Kr indicates that the high-$j$ proton orbital $g_{9/2}$ intrudes into the $pf$-shell region near the Fermi surface. In the other case, the $0^+-8^+$ states in $^{78}$Se are mainly characterized by the $\pi(f_{5/2}^4p_{3/2}^2)\otimes \nu(g_{9/2}^6)$ configuration and proton excitations in the $g_{9/2}$ orbital play less significant role. However, for the $10^+$ state, a proton-pair excitation in the $g_{9/2}$ orbital is found, with major components from the [$\pi(f_{5/2}^3p_{3/2}^1g_{9/2}^2)\otimes \nu(g_{9/2}^6) \sim 21.0\%$], [$\pi(f_{5/2}^2p_{3/2}^2g_{9/2}^2)\otimes \nu(g_{9/2}^6) \sim 19.7\%$], and [$\pi(f_{5/2}^4g_{9/2}^2)\otimes \nu(g_{9/2}^6) \sim 19.3\%$] configurations.

The kinematic moment of inertia (I$/\hbar^2$) is an important quantity to describe the rotational behavior of a band. In Fig. \ref{kinematic_inertia}, we have shown the variation of theoretical and experimental kinematic moments of inertia in $^{78}$Kr and $^{78}$Se as a function of yrast positive-parity spins. The relation I/$\hbar^2 = (2J - 1)/E_{\gamma}(J \rightarrow J-2)$ is used, where $E_{\gamma}$ represents the energy difference between two consecutive states in the g.s. band, which are at the spin difference of $\Delta J=2$. Our theoretical results show quite good agreement with the experimental data, particularly for $^{78}$Kr, compared to the previous study \cite{Liu}.

 As reported in Table \ref{BE2}, the calculated $B(E2)$ strengths for both the $^{78}$Kr and $^{78}$Se are slightly lower but consistent with the experimental data \cite{NNDC_NUDAT}. Although if we incorporate the neutron orbitals $0g_{7/2}1d2s$ in our model space with $1p-1h$ excitation across $N=50$, then it is possible to improve the $B(E2)$ strengths, but these orbitals cannot be included in the current study due to the computational limitations for the 2$\nu$ECEC NME calculation with higher $1^+$ state energies in the $^{78}$Br. Nevertheless, the shell-model predicted energy spectra, kinematic moment of inertia, and $B(E2)$ strengths show quite reasonable agreement with the experimental data for both nuclei. Thus, the present effective interaction can be reliably used to calculate the 2$\nu$ECEC NME in this study.

\vspace{1mm}

\subsection{Level density of $1^+$ states}

The Bethe formula for level density \cite{Bethe}, incorporating spin dependence within the BFM, can be expressed as \cite{Gilbert}

\begin{equation}
    \rho(U,J) = \rho(U) \cdot f(J),
\end{equation}

where

\begin{equation}
    \rho(U) = \frac{1}{12\sqrt{2}\sigma} \frac{e^{2\sqrt{aU}}}{a^{1/4}U^{5/4}},
\end{equation}

\vspace{1mm}

and

\begin{equation}
    f(J) = \frac{(2J+1)e^{-\left(J + \frac{1}{2}\right)^2 / 2\sigma^2}}{2\sigma^2}.
\end{equation}

In this formulation, the spin cutoff parameter \cite{Gilbert} is described as $\sigma^2 = 0.0888 A^{2/3} \sqrt{a(E - \delta)}$. The level density parameter $a$ is defined as $a = (A/16)$ MeV$^{-1}$ \cite{Gross}, and $U = (E - \delta)$ \cite{Koning}, where $E$ is the excitation energy for a given spin state $J$, and $A$ represents the atomic mass. The back-shifted parameter $\delta$ (in MeV) is determined by the following expression \cite{Rauscher}

\begin{equation}
 		\delta =
 		\left\{
 		\begin{array}{ll}
 			12/\sqrt{A}, & {\rm for ~ even-even ~ nuclei} \\
 			0, & {\rm for ~ odd-{\it A} ~ nuclei} \\
 			-12/\sqrt{A}, & {\rm for ~ odd-odd ~ nuclei}.
 		\end{array}
 		\right.
 	\end{equation}

By accounting for equiparity distribution, the level density can be represented as follows \cite{Koning}

\begin{equation}
    \rho(U,J,\pi) = \frac{1}{2} \rho(U) \cdot f(J).
    \label{eq8}
\end{equation}

\begin{figure}
	\centering
	\includegraphics[width=15.1cm]{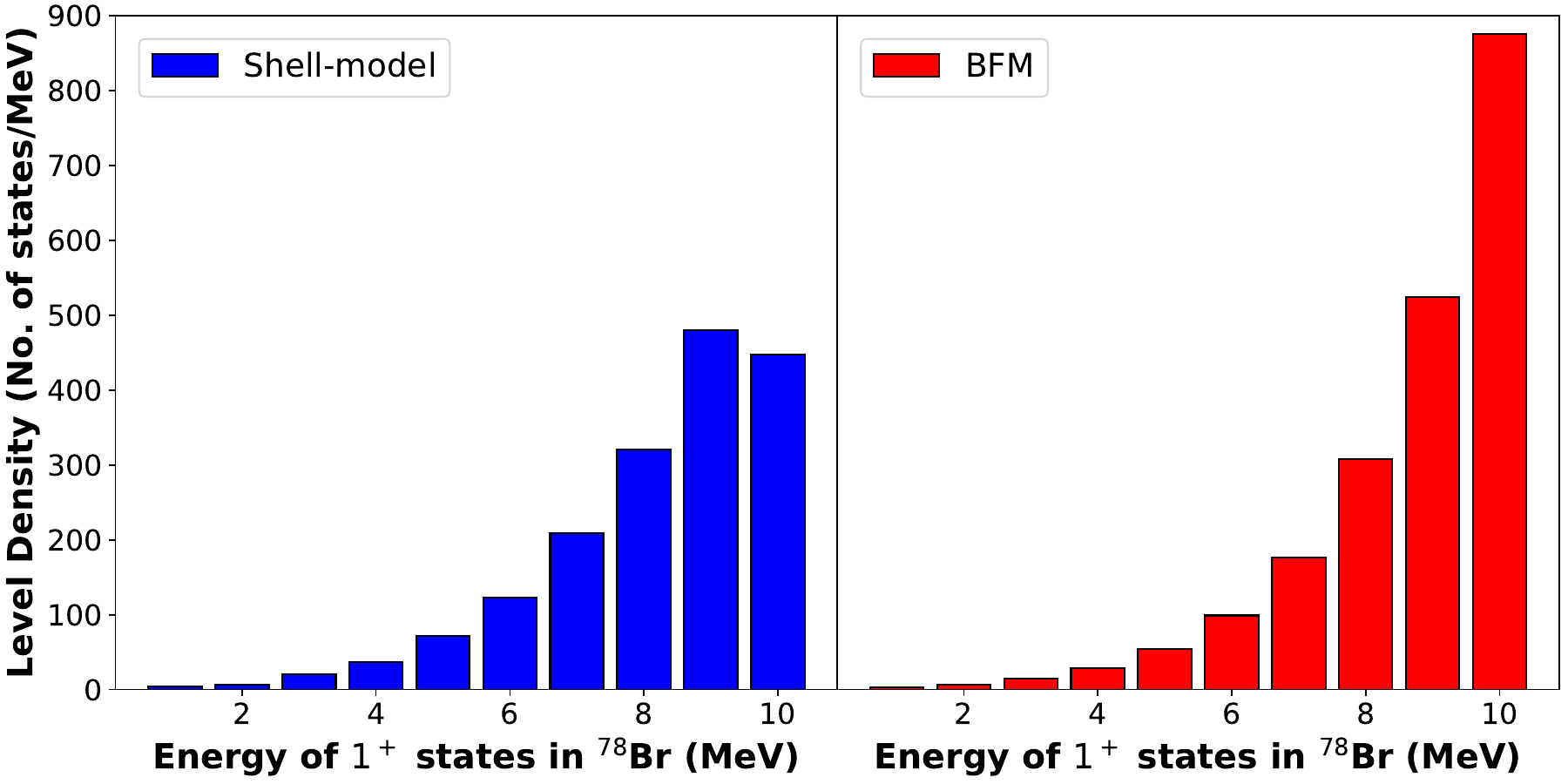}
	\caption{\label{level_density} Comparison between the shell-model and BFM calculated level-densities of $1^+$ states in $^{78}$Br.}
\end{figure}

Fig. \ref{level_density} depicts the comparison of the shell-model calculated level density of $1^+$ states in $^{78}$Br with the BFM predicted level density for odd-odd nuclei with $A=78$ (using Eq. (\ref{eq8})). It is clear from Fig. \ref{level_density} that the shell-model level densities of $1^+$ states in $^{78}$Br are consistent with the BFM predictions up to approximately 9 MeV. Therefore, the cumulative NME contribution for the $2\nu$ECEC process in $^{78}$Kr is expected to be well captured by shell-model calculations up to this energy. Beyond 9 MeV, some contributions to the cumulative NME may be missed by the shell-model calculations but will be significant in the final value of $|M_{2\nu}|$ only if the cumulative NME has not yet reached the saturation level. The inclusion of further neutron and proton orbitals in the model space can extend the energy range to sustain the level density of the $1^+$ state in $^{78}$Br by shell-model with more advanced computational facilities, which is the future work.

\begin{figure}
	\centering
	\includegraphics[width=15cm]{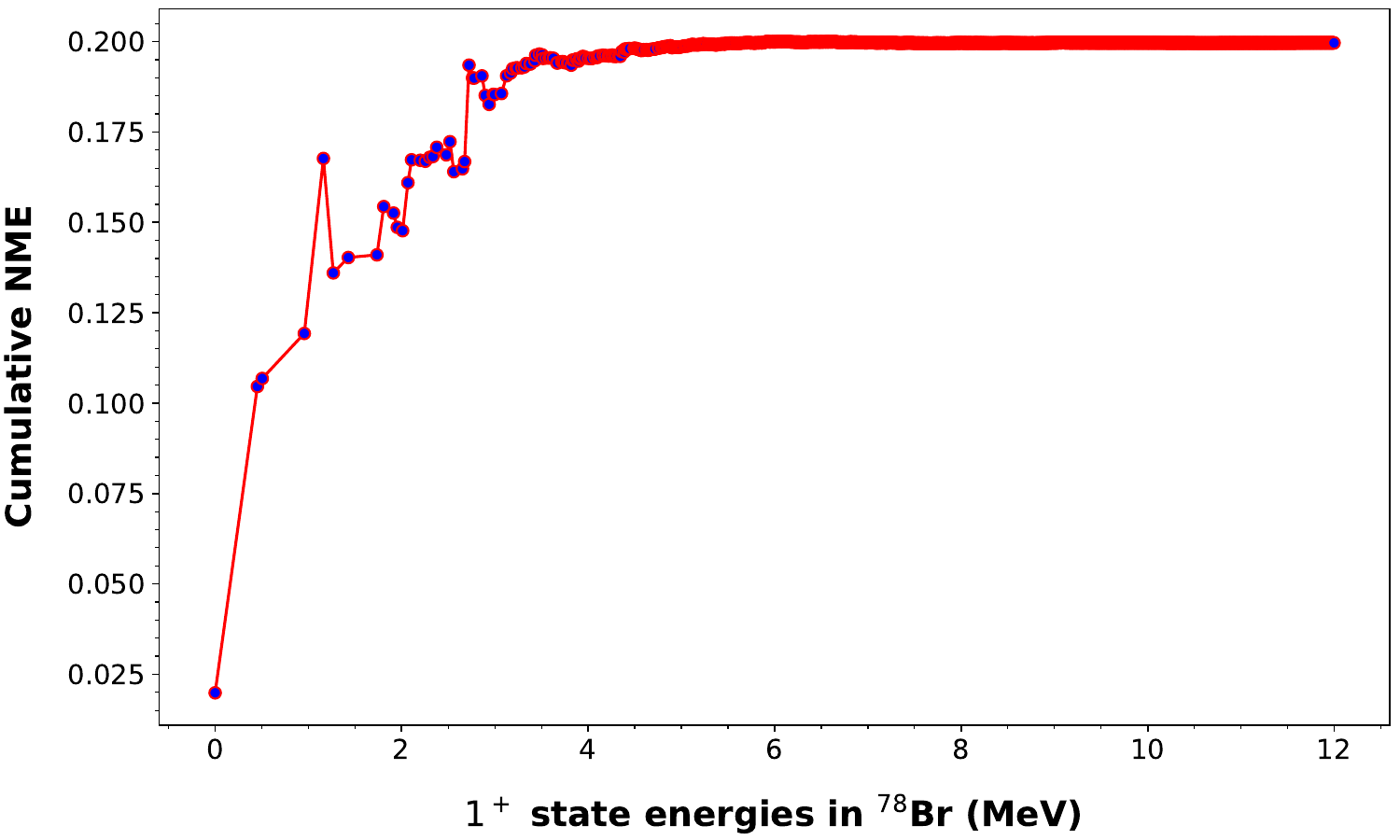}
	\caption{\label{NME_vary} Variation of the cumulative NME with respect to the $1^+$ state energies in the intermediate nucleus $^{78}$Br.}
\end{figure}

\subsection{Variation of cumulative NME and half-life}

              \begin{table}
			\centering
			\caption{Shell-model calculated $2\nu$ECEC NMEs and the extracted half-lives.}
			\begin{tabular}{cccccc}
				\hline\hline
				Nucleus	& $|M_{2\nu}|$ & $G_{2\nu}^{\rm ECEC}$ (yr$^{-1}$) \cite{Nitescu}  & $g_{A}^{\rm eff}$  & Calculated $T_{1/2}^{2\nu}$ (yr) &  \makecell{Experimental \\ value of $T_{1/2}^{2\nu}$ (yr)}  \\
				\hline
				&  &   & & &  \\
				
				$^{78}$Kr	& 0.1997  & $520.518\times 10^{-24}$ & \makecell{1.27- \\ 1.00} & (1.85-4.82)$\times 10^{22}$  & \makecell{$[9.2^{+5.5}_{-2.6}({\rm stat}) \pm $ \\ 1.3({\rm syst})$]\times 10^{21}$} \cite{Gavrilyuk}  \\

                    & & & & & \makecell{$[1.9^{+1.3}_{-0.7}({\rm stat}) \pm $ \\  $ 0.3 ({\rm syst})]\times 10^{22}$} \cite{Ratkevich} \\
				
				\hline\hline	
				
			\end{tabular}
			\label{half-life}
		\end{table}

Here, we analyze the variation in the shell-model predicted cumulative NME for the $2\nu$ECEC process in $^{78}$Kr and compare the half-life, calculated from the final NME (using Eq. \ref{eq1}) with the available experimental data. Accurate determination of the NME is essential for the estimation of precise half-life. Fig. \ref{NME_vary} depicts the variation of cumulative NME as a function of the $1^+$ state energies in the $^{78}$Br, up to the saturation level. In the present calculation, the excitation energies of the $1^+$ states in $^{78}$Br are shifted so that the lowest $1^+$ state comes at the experimental energy of 0 MeV. The $Q_{\rm ECEC}$ value is taken from the Ref. \cite{Nitescu}. The lowest $1^+$ state in $^{78}$Br contributes 9.95\% to the final NME. After several significant constructive and destructive contributions, likely within the 0-5 MeV range, the cumulative NME begins to saturate approximately near the 8 MeV, with a final value of 0.1997. It is important to mention that since the cumulative NME nearly saturates before the energy range where our model space becomes less efficient in sustaining the level density of the $1^+$ states in $^{78}$Br. Thus, it has minimal effect on the final NME value. In Fig. \ref{NME_vary}, the cumulative NME is presented up to 12 MeV, accounting for 2320 $1^+$ states in $^{78}$Br, as it has already saturated at the lower energies.

The estimated half-life compared to the experimental data is presented in Table \ref{half-life}. The experimental half-lives reported in Table \ref{half-life} are corresponding to the electron captures from the K-shell pairs. Therefore, we use the $G^{\rm ECEC}_{2\nu}=661.9 \times 0.7864 \times 10^{-24}$ yr$^{-1}$ \cite{Nitescu} in our calculations, which is appropriate for the 2$\nu$KK mode. We have considered a range of $g_A^{\rm eff}$ from $g_A^{\rm free}=1.27$ to a quenched value of $g_A^{\rm quen}=1.00$. The calculated half-life using the shell-model predicted NME shows quite reasonable agreement with the experimental results \cite{Gavrilyuk, Ratkevich}. Previously, Rumyantsev $et$ $al.$ \cite{Rumyantsev} calculated the $2\nu$ECEC NME for $^{78}$Kr as 0.0146 using the BCS and the pair-vibration models, a small value that contributed to a larger half-life prediction compared to the experimental data. However, recent half-life measurement has yielded an NME for the $2\nu$ECEC process in $^{78}$Kr of $0.318^{+0.100}_{-0.073}$ \cite{Ratkevich, Nitescu}, which is approximately 1.6 times larger than the NME obtained in the present study. In the future, it may be possible to achieve more accurate NME from theoretical calculations with a larger model space using advanced computational resources.

\subsection{GT transitions from the lowest $1^+$ state of the $^{78}$Br}

Experimentally, the GT transitions from the lowest $1^+$ state in the $^{78}$Br can occur through both the EC$+\beta^+$ and $\beta^-$-decay channels. However, only limits on the branching ratios ($B.R.$) for the $^{78}{\rm Br}(1^+_1) \rightarrow ^{78}$Se$(0^+, 2^+)$ and $^{78}{\rm Br}(1^+_1) \rightarrow ^{78}$Kr$(2^+_1)$ transitions are known, with values of $\ge 99.99$\% and $\le 0.01$\%, respectively \cite{NNDC, Hinrichsen}. As reported in Ref. \cite{Hinrichsen}, the dominant contribution (99.7\%) to the EC+$\beta^+$ decay path arises from the $^{78}{\rm Br}(1^+_1) \rightarrow ^{78}$Se$(0^+_{\rm g.s.})$ and $^{78}{\rm Br}(1^+_1) \rightarrow ^{78}$Se$(2^+_1)$ transitions. By considering only these two transitions for the EC+$\beta^+$ channel and the $^{78}{\rm Br}(1^+_1) \rightarrow ^{78}$Kr$(2^+_1)$ transition for the $\beta^-$-channel in our shell-model calculations, we obtain a total half-life of 56.34 min for the $1^+_1$ state in $^{78}$Br. This value is of the same order as the experimental half-life (6.45(4) min \cite{NNDC}) but approximately 8.7 times larger. The partial half-lives for the $^{78}$Br$(1^+_1) \rightarrow ^{78}$Se$(0^+_{\rm g.s.})$, $^{78}$Br$(1^+_1) \rightarrow ^{78}$Se$(2^+_1)$, and $^{78}$Br$(1^+_1) \rightarrow ^{78}$Kr$(2^+_1)$ transitions are calculated as 58.34, 1863.32, and 13394.14 min, respectively. The branching ratio for the combined transitions $^{78}$Br$(1^+_1) \rightarrow ^{78}$Se$(0^+_{\rm g.s.})$ (96.56\%) and $^{78}$Br$(1^+_1) \rightarrow ^{78}$Se$(2^+_1)$ (3.02\%) is 99.58\%. While, the remaining 0.42\% is attributed to the $^{78}$Br$(1^+_1) \rightarrow ^{78}$Kr$(2^+_1)$ transition. These $B.R.$ are close to the experimental limits. An analogous situation has also been found in the case of $\beta^-\beta^-$-decay of $^{100}$Mo \cite{NNDC, Sjue}. The estimated partial half-lives are corresponding to the $g_{\rm A}^{\rm eff} = 1.00$. Although the $B.R.$ remain unaffected by changing the $g_{\rm A}^{\rm eff}$. The total half-life for the $1^+_1$ state in $^{78}$Br is calculated as 34.93 min with $g_{\rm A}^{\rm free} = 1.27$, which aligns more closely with the experimental value than corresponding to the $g_{\rm A}^{\rm eff} = 1.00$. This improvement is also reflected in the calculated half-life for the $2\nu$ECEC process, as shown in Table \ref{half-life}. Therefore, our nuclear model, which accurately estimates the half-life of the $1^+_1$ state in $^{78}$Br, is also likely to yield a precise prediction of the half-life for the $2\nu$ECEC process in $^{78}$Kr. These analyses support the dominance of this $1^+_1$ state in contributing to the $2\nu$ECEC process.

The partial half-lives are calculated using the formula $t_{1/2}^{(n)}=\kappa/(f_0 B_{\rm GT}^{(n)})$ \cite{Vikas}. Here, $n$ stands for the different GT transitions; $\kappa$ denotes the universal constant, which equals to 6289 s \cite{Patrignani}; $B_{\rm GT}^{(n)}$ denotes the GT reduced transition probability and given in terms of reduced GT matrix elements $\langle J_{f}||\sigma\tau^{\pm}||J_{i} \rangle$ as follows \cite{Suh2007}

\begin{equation}
    B_{\rm GT}=\frac{(g_{\rm A}^{\rm eff})^2}{2J_i+1}|\langle J_{f}||\sigma\tau^{\pm}||J_{i}\rangle|^2,
\end{equation}

where $J_i$($J_f$) is spin of the initial (final) state. The $f_0$ is the phase-space factor and can be written for $\beta^{\pm}$-decay as follows \cite{Haaranen}

        \begin{equation}
            f_0^{(\pm)}=\int^{E_0}_{1} F_0(\mp Z_f,\epsilon)p\epsilon(E_0-\epsilon)^2d\epsilon.
        \end{equation}

Here, $F_0$ denotes the Fermi function; $Z_f$ stands for the atomic number of the final nucleus. The other parameters can be given by

        \begin{equation}
            \epsilon=\frac{E_e}{m_ec^2},~ E_0=\frac{E_i-E_f}{m_ec^2},~ p=\sqrt{\epsilon^2-1}=\frac{p_ec}{m_ec^2},
        \end{equation}

where, $E_e$ represents the total energy of the emitted positron/electron and $E_i(E_f)$ denotes the energy of the initial (final) nuclear state.

The phase-space factor for electron capture (EC) can be written as \cite{Suh2007}

        \begin{equation}
            f_0^{(EC)}=2\pi (\alpha Z_i)^3 (\epsilon_0 +E_0)^2,
        \end{equation}

        where, $\epsilon_0=1-[(\alpha Z_i)^2]/2$, and the fine-structure constant, $\alpha\approx 1/137$; $Z_i$ denotes the atomic number of initial nucleus.

Finally, we have calculated the total half-life and $B.R.$ using the following expressions \cite{Suh2007}

\begin{equation}
\frac{1}{T_{1/2}^{\rm total}}= \sum_j \frac{1}{t_{1/2}^{(n)}}; ~~~~~~ B.R.^{(n)}=T_{1/2}^{\rm total}/t_{1/2}^{(n)}.
\end{equation}

Here, it is important to mention that when we consider only the single transitions $^{78}$Br$(1^+_1) \rightarrow ^{78}$Se$(0^+_{\rm g.s.})$ for the EC$+\beta^+$ decay and $^{78}$Br$(1^+_1) \rightarrow ^{78}$Kr$(0^+_{\rm g.s.})$ for the $\beta^-$-decay, the resulting partial half-lives corresponding to $g_{\rm A}^{\rm free} = 1.27$ are 36.17 min for the EC$+\beta^+$ channel and 254.21 min for the $\beta^-$-channel. In this case, the calculated $B.R.$ (87.54\% for the EC$+\beta^+$ channel and 12.46\% for the $\beta^-$-channel) show some deviations from the experimental limits, which may pose a smaller effect on the accuracy of the calculated NME.

\section{Summary and conclusion} \label{section4}

In this study, we have conducted large-scale shell-model calculations to compute a precise NME for the $2\nu$ECEC process in the $^{78}$Kr, utilizing the GWBXG effective interaction. To validate the efficiency of this interaction, we first calculated the spectroscopic properties, such as energy spectra, kinematic moment of inertia, and reduced quadrupole transition probabilities for the parent ($^{78}$Kr) and grand-daughter ($^{78}$Se) nuclei, as well as the level density of $1^+$ states in the intermediate nucleus $^{78}$Br. Our results exhibit quite remarkable agreement with the experimental data, especially in the low-lying energy spectra and kinematic moment of inertia. The model space effectively captures the level density of the $1^+$ states in $^{78}$Br up to 9 MeV, though some deviation occurs when compared to the predicted level density from the BFM at higher energies. Analyzing the cumulative NME as a function of $1^+$ state energies in the $^{78}$Br revealed that the cumulative NMEs start to saturate around 8 MeV, indicating that our model space successfully incorporates the significant contributions to the final cumulative NME. Based on the computed NME, we have estimated the half-life of $^{78}$Kr for the $2\nu$ECEC process corresponding to the electron captures from the K-shell pairs, which shows quite reasonable agreement with the experimental data. The role of GT transitions for the $2\nu$ECEC process in $^{78}$Kr from the lowest $1^+$ state of $^{78}$Br via both the EC$+\beta^+$ and $\beta^-$-channels is also briefly discussed. We will explore the $2\nu$ECEC process in the $^{78}$Kr and other candidates with larger model spaces to get more accurate NME in the future once the new computational facilities allow us to do so.

\section*{{Acknowledgments}}

We are thankful for the financial support provided by MHRD, the Government of India, and SERB (India), CRG/2022/005167. We would like to acknowledge the National Supercomputing Mission (NSM) for providing computing resources of ‘PARAM Ganga’ at the IIT Roorkee, implemented by C-DAC and supported by the Ministry of Electronics and Information Technology (MeitY) and Department of Science and Technology (DST), Government of India.


\end{document}